\documentclass[journal,10pt,a4paper,twocolumn,twoside]{IEEEtran}

\IEEEoverridecommandlockouts
\usepackage{adjustbox}
\usepackage{comment}
\usepackage{cite}
\usepackage{physics}
\usepackage{caption} 
\usepackage{amsmath,amssymb,amsfonts,amsthm}
\usepackage{mathtools}
\usepackage{algorithmic}
\usepackage{graphicx}
\usepackage{textcomp}
\usepackage{tikz}
\usetikzlibrary{automata, positioning, arrows,calc}
\usetikzlibrary{quantikz}
\usepackage{algorithm}
\usepackage{algorithmic}
\usepackage{float}
\usepackage{subcaption}
\usepackage[T1]{fontenc}
\usepackage{hyperref}
\usepackage{multirow}
\usepackage{booktabs}
\usepackage{multicol}
\usetikzlibrary{decorations.pathreplacing}

\definecolor{myLightGray}{RGB}{191,191,191}
\definecolor{myGray}{RGB}{160,160,160}
\definecolor{myDarkGray}{RGB}{144,144,144}
\definecolor{myDarkRed}{RGB}{167,114,115}
\definecolor{myRed}{RGB}{255,58,70}
\definecolor{myGreen}{RGB}{0,255,71}

\theoremstyle{remark}
\theoremstyle{definition}

\newtheorem*{rem}{Remark}

\begin{document}

\def\myvdots{\ \vdots\ }

\title{Quantum Internet Addressing}
\author{Angela~Sara~Cacciapuoti,~\IEEEmembership{Senior~Member,~IEEE}, Jessica Illiano, Marcello~Caleffi,~\IEEEmembership{Senior~Member,~IEEE}
    \thanks{A.S. Cacciapuoti, J. Illiano, M. Caleffi, are with the \href{www.quantuminternet.it}{www.QuantumInternet.it} research group, \textit{FLY: Future Communications Laboratory}, University of Naples Federico II, Naples, 80125 Italy. E-mail: \href{mailto:angelasara.cacciapuoti@unina.it}{angelasara.cacciapuoti@unina.it}, \href{mailto:jessica.illiano@unina.it}{jessica.illiano@unina.it},
    \href{mailto:marcello.caleffi@unina.it}{marcello.caleffi@unina.it}. Web: \href{http://www.quantuminternet.it}{www.quantuminternet.it}.}
%    \thanks{A.S. Cacciapuoti and M. Caleffi are also with the Laboratorio Nazionale di Comunicazioni Multimediali, National Inter-University Consortium for Telecommunications (CNIT), Naples, 80126, Italy.}
    \thanks{Angela Sara Cacciapuoti acknowledges PNRR MUR NQSTI-PE00000023, Marcello Caleffi acknowledges PNRR MUR project RESTART-PE00000001.}
}

\newcommand{\eqdef}{\stackrel{\triangle}{=}}

\maketitle

\begin{abstract}
The design of the Quantum Internet protocol stack is at its infancy and early-stage conceptualization. And different heterogeneous proposals are currently available in the literature. The underlying assumption of the existing proposals is that they implicitly mimic classical Internet Protocol design principles: ``\textit{A name indicates what we seek. An address indicates where it is. A route indicates how to get there}''. Hence the network nodes are labeled with classical addresses, constituted by classical bits, and these labels aim at reflecting the node location within the network topology. In this paper, we argue that this twofold assumption of \textit{classical} and \textit{location-aware} addressing constitutes a restricting design option, which prevents to scale the quantumness to the network functionalities, beyond simple information encoding/decoding. On the contrary, by embracing quantumness within the node addresses, quantum principles and phenomena could be exploited for enabling a quantum native functioning of the entire communication network. This will unleash the ultimate vision and capabilities of the Quantum Internet.
%The design of the Quantum Internet cannot be limited to simply replace classical information encoding/decoding with some equivalent quantum functionality. Indeed, the intrinsic dissimilarities between classical and quantum information requires a paradigm shift embracing the design of all the network functionalities. In the following, we provide the rationale for this vision by focusing on a specific network functionality -- namely, quantum addressing -- hoping to start a discussion and to pave the way toward a holistic design of the Quantum Internet protocol stack able to harness the peculiarities of quantum information and quantum entanglement. More into details, usual Quantum Internet design restricts the node addressing to be classical. This implies the impossibility to scale the quantumness to the network functionalities, beyond simple information encoding/deconding. In this paper we argue that, by embracing quantumness within the node addresses, quantum principles and phenomena could be exploited for enabling a native quantum functioning of the entire communication network. 
\end{abstract}


% ------------------------------------------------------
% Sec. I
% ------------------------------------------------------
\section{Introduction}
\label{Sec:1}

The Quantum Internet is envisioned as the final stage of the quantum revolution, opening fundamentally new communications and computing capabilities beyond quantum cryptography \cite{KozWehVan-22}. 

These unparalleled functionalities have the potential of radically changing the world in which we live in ways we cannot imagine yet. As a matter of fact, a preliminary set of specifications for the Quantum Internet has already being drafted, with several experimental and standardization efforts, ranging from IETF with the seminal ``architectural principles'' RFC \cite{KozWehVan-22}, to ITU, IEEE, GSMA, and ETSI.

In this vibrant context, the state-of-art related to the design of the Quantum Internet protocol stack is at its infancy and early-stage conceptualization \cite{IllCalMan-22,CaccIllKou-22}. Hence, we are still very far from having a complete and univocal protocol model, as we have for classical Internet. And different heterogeneous proposals are currently available in the literature\cite{IllCalMan-22},  \cite{LiXueChe-23}. 

Indeed, the underlying hypothesis of the existing proposals is to implicitly mimic classical Internet Protocol (IP) design principles: ``\textit{A name indicates what we seek. An address indicates where it is. A route indicates how to get there}''   as declared in RFC791 . Hence, network nodes are implicitly labeled with classical addresses, constituted by classical bits, and these labels aim at reflecting the node location within the network topology.

In this paper, we argue that this twofold assumption of \textit{classical} and \textit{location-aware} addressing constitutes a restricting design option, which prevents to scale the quantumness to the network functionalities, beyond simple information encoding/decoding.

%Specifically, existing literature allows the message -- i.e., the packet \textit{payload} borrowing IP terminology -- to be quantum in order to harness capabilities with no classical counterpart. Yet, control information such as network addresses is restricted to be classical. Clearly, this design choice fails in scaling quantumness to the network functionalities, beyond ``simple'' information encoding/decoding.

%This is a drawback in the Quantum Internet, since classical IP addresses fail in scaling quantumness beyond information encoding/decoding.
%and hence fail in capturing the dynamic nature of entanglement-based connectivity, as clarified in the following.

Conversely, by \textit{embracing} quantumness within the network functionalities, quantum principles and phenomena could be exploited for enabling a native quantum functioning of the entire communication network. This can be regarded as an additional level of Internet quantization, where the original level was to quantize the messages delivered by the network, while the second level is to quantize the network functionalities.

To this aim, a \textit{quantum addressing} is a mandatory pre-requisite for any network functionality design, and in the following we will focus on it, as archetypal case study capable of providing the reader with an overview of the potentialities offered by a native quantum network functioning. 
 
Specifically, the main contributions of the paper are summarized as follows:
\begin{itemize}
    \item we discuss the key drawbacks arising by adopting classical, location-aware addressing within the Quantum Internet;
    \item we propose the novel \textit{quantum addressing} functionality for the Quantum Internet, 
    \item we discuss how, by embracing quantumness within the node addresses, it is possible to unleash the advantages enabled by quantum  propagation of information carriers though the concept of \textit{quantum paths};
    \item we discuss the impact of quantum addressing on the design of the Quantum Internet, and we propose a toy-model of a quantum addressing scheme able to overcome the limitations of classical location-aware addressing schemes through link augmentation;
    \item we provide the reader with insights on future research directions and open issues to be addressed towards the ultimate vision of the Quantum Internet.
\end{itemize}

% ------------------------------------------------------
% Sec. II
% ------------------------------------------------------
\section{Background: Classical Internet Addressing}
\label{Sec:2}

\begin{figure*}
    \centering
    \includegraphics[width=0.8\linewidth]{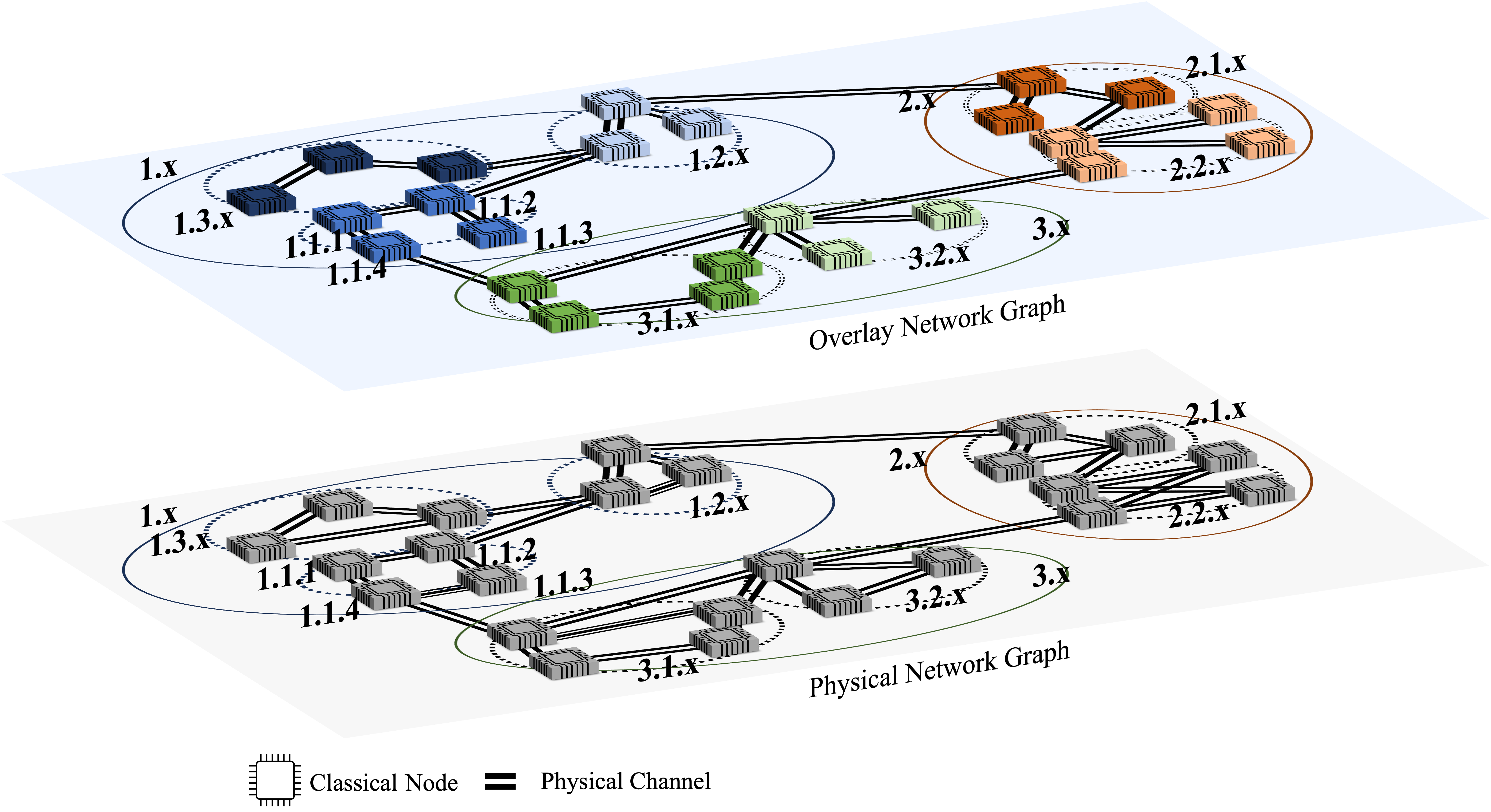}
    \caption{Classical Hierarchical Addressing. Internet routing scalability is achieved by hierarchical clustering the network nodes, so that routing tables keep only one entry for all the nodes in each cluster level. From a topological perspective, this implies that path discovery is not performed through the entire physical network, but rather through the overlay \textit{routing} network implemented according to the incomplete topological information stored within the routing tables. }
    
    \label{fig:01}
    \hrulefill
\end{figure*}
Since Kleinrock's seminal work \cite{KleKam-77} dated over forty years ago, classical Internet routing has pursued scalability mainly through clustering.

More into details, it is well known that maintaining complete topological knowledge about the network topology at each node, through one routing table entry for each destination, becomes quickly prohibitive both in terms of \textit{storage} and \textit{update} cost, as the number of network nodes grows.

Hence, the \textit{key design principle} behind classical Internet routing has been the wisely selection of the partial topological information to be stored at each node, by reducing substantially the size of the routing tables. For this, topological details about remote portions of the network are discarded. Hence, at each node, (almost) complete topological information should be available for destinations that are close\footnote{Close according to some meaningful metric from a topological perspective, with the representative example constituted by hop-count.} to the node, whereas less information should be maintained for destinations far away. And the further the destination is, the less the information is.

This design principle is achieved through \textit{hierarchical clustering} of the network: nearby nodes are grouped into clusters, clusters into super-clusters, and so on in a bottom-up fashion with multiple levels of hierarchy among clusters. Routing tables are thus organized so that they keep only one entry for all the nodes in each cluster level and, if the cardinality of the clusters grows exponentially as the level increases (i.e., $\mathcal{O}(2^k)$ nodes in a $k$-level cluster), the number of routing entries scales logarithmically with the network size. Almost all the proposals for classical Internet trying to address routing table scalability -- included the ones used nowadays in the form of  Classless Inter-Domain Routing (CIDR)    for inter-domain routing or OSPF/ISIS (Open Shortest Path First / Intermediate System to Intermediate System) areas for the intra-domain routing -- are based, explicitly or implicitly, on the hierarchical routing principle  \cite{KriClaFal-07}.

It must be noted though that classical Internet topology is not a static hierarchical topology per-se, such as the one exhibited by regular static graphs like trees or grids. Rather, it is a dynamic topology exhibiting \textit{scale-free} characteristics.

This implies that the topological information reduction, rather than based on some peculiar characteristics of the underlying physical graph, is fundamentally obtained by embedding some topological information within node labels. Thus, node labels cannot be arbitrary identities -- i.e., \textit{flat} addresses such as IEEE 802 Medium Access Control (MAC) ones --  but they must somehow reflect the node location within the network topology, as it happens with IP addresses by design.

Unfortunately, location-aware addressing such as IP one doesn't come for free. As instance, it requires extensive assignment planning and management, as well as additional network functionalities, with   Domain Name System (DNS)   as pivotal example, for mapping univocal node identities (i.e., names in IP terminology) to node addresses. Furthermore, the reduction of the topological information stored at each node implies a sub-optimality of the path discovery process, regardless of the particulars of the adopted routing protocol. Indeed, packets can be forwarded through longer routes.

Overall, from a network perspective, Internet routing scalability is achieved through \textbf{topological depletion}. Path discovery is not performed through the entire physical network, rather it is performed through the overlay \textit{routing} network implemented according to the incomplete topological information stored within the routing tables. Hence, the overlay network is built on top of the underlying physical network by incorporating only a subset of the links available within the physical graph, as represented in Figure~\ref{fig:01}. This topological depletion allows to reduce Internet physical graph structure to some regular graph, such as a tree, which in turn is strongly influenced by the underlying network characteristics.

% ------------------------------------------------------
% Sec. III
% ------------------------------------------------------
\section{Quantum Addressing}
\label{Sec:3}

\begin{figure*}
    \centering
    \includegraphics[width=0.8\linewidth]{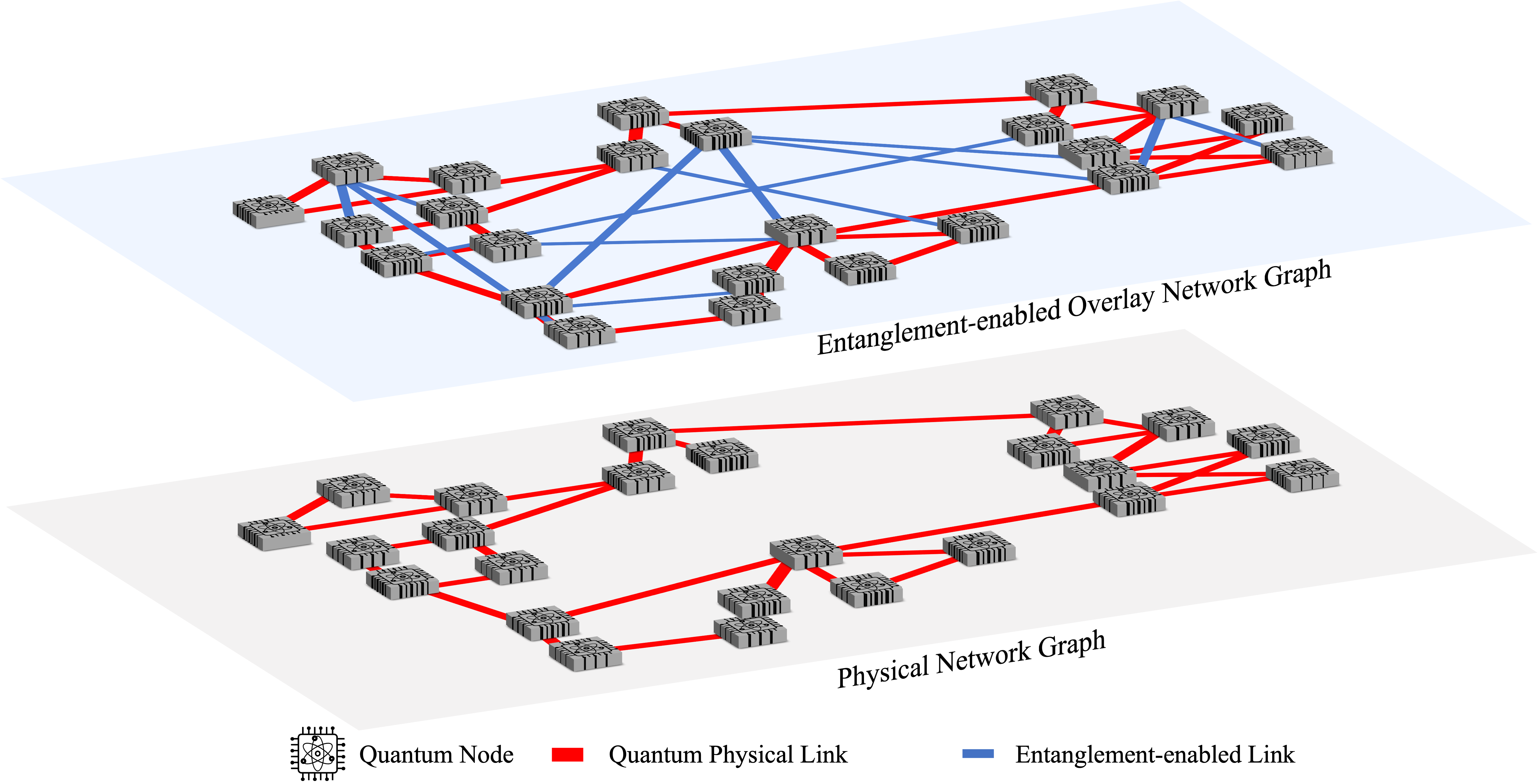}
    \caption{Entanglement-enabled Connectivity: \textit{Physical Network Graph} versus \textit{Entanglement-enabled Overlay Network Graph}.}
    \label{fig:02}
    \hrulefill
\end{figure*}

As discussed in Section~\ref{Sec:1}, quantum addressing -- as the quantum equivalent of the univocal network addressing provided by IP and its consequences on routing within the Quantum Internet -- is yet an unexplored research domain. A notable exception is \cite{RamPirDur-21}, where quantumness is exploited for enabling quantum networks to perform different tasks and to address other devices in a coherent fashion through control quantum registers.   It must be noted that the quantum addressing is not envisioned as a substitute for the classical addressing. Indeed, a classical address is needed at the quantum network nodes for classical communications and classical signaling which are mandatory for any quantum communication protocol.

In this paper, for the first time to the best of our knowledge, we discuss key drawbacks arising by adopting \textit{classical}, \textit{location-aware} addressing within the Quantum Internet, namely, i) failing in modeling the peculiarities of entanglement-enabled connectivity, ii) failing in embracing the unique propagation characteristics of quantum information carriers, and iii) failing in modeling the very fundamental goal of Quantum Internet routing. 

% ------------------------------------------------------
\subsection{Entanglement-Enabled Connectivity}
\label{Sec:3.1}

From a network perspective, entanglement enables a new and richer form of connectivity, with respect to classical networks \cite{IllCalMan-22}.

Specifically, once an entangled state -- say an Einstein-Podolsky-Rosen (EPR) pair for the sake of exemplification -- has been shared between two nodes, a qubit can be ``transmitted''  via quantum teleportation, regardless of the instantaneous conditions of the physical quantum link connecting the two nodes. Remarkably, qubit transmission is still possible even when there is no longer a quantum link connecting the nodes together. In this sense, entanglement enables a new form of connectivity, referred to as \textit{entanglement-enabled connectivity}, which differs from classical Internet connectivity in that: i) it exhibits a weaker dependency on the underlying physical communication link, and ii) it exhibits unconventional temporal dynamics, since entanglement is depleted once used.

Furthermore, entanglement can be swapped and, hence, it is possible to dynamically, namely, at run-time, change the identities of the entangled nodes. Hence, entanglement redefines the very same concept of topological \textit{neighborhood}\footnote{It is worthwhile to underline that neighborhood is a crucial concept in classical Internet routing, where the store-and-forward paradigm exploits neighbor nodes for delivering packets to remote nodes.}, with no counterpart in the classical world \cite{IllCalMan-22}. Accordingly, entanglement enables half-duplex unicast links between any pairs of nodes, regardless of their relative positions within the underlying physical network topology.  In other words, any pair of nodes can be neighbor as long as they share entanglement.

Additionally, entanglement is not limited to EPR pairs. Indeed, when it comes to multipartite entanglement, the dynamic nature of the entanglement-based connectivity becomes even more evident. As instance, by distributing an $n$-qubit Greenberger–Horne–Zeilinger (GHZ) state among $n$ network nodes, an EPR pair can be distributively extracted by any pair of nodes, with the identities of the entangled nodes chosen at run-time. Hence, even if all the $n$ nodes are possible neighbor nodes, only two out of $n$ can actually exploit the entanglement to create a half-duplex unicast link.

From the above, it becomes evident that any addressing scheme for the Quantum Internet, aiming at achieving routing scalability, cannot resort to classical node identities reflecting node location within the physical network topology, as it happens with classical IP addresses. Rather, it should aim at properly capturing and tracking the rich, dynamic nature of entanglement-enabled connectivity. As a matter of fact, entanglement-enabled connectivity should not only be captured by the quantum addressing. But, it should be properly \textit{engineered} for improving the routing process, as further discussed in Section~\ref{Sec:4}.

Specifically, as described in Section~\ref{Sec:2}, hierarchical routing achieve scalability through incomplete topological information. This is equivalent to build an overlay routing network with special graph properties through topological depletion, by storing within the routing tables only a subset of the forwarding possibilities offered by the physical neighbors.

Conversely, entanglement-enabled connectivity allows to augment the neighbor set, by creating ``additional'' links toward remote nodes through entanglement swapping. Hence, it enables the possibility to build and to engineer an \textit{overlay entangled network} where the network graph properties needed by the routing process are obtained through \textit{topological augmentation}, rather then topological depletion, as depicted in Figure~\ref{fig:02}. This possibility will be further discussed in Section~\ref{Sec:4}.

\begin{rem}
It is important to highlight that, although overlay networks enabled by entanglement share some similarities  with classical virtual overlay networks -- such as those arising, as instance, with peer-to-peer (P2P) systems -- entanglement-enabled connectivity unlocks characteristics with no classical counterpart, as discussed in the following.
\end{rem}

Classical overlay networks aim at form virtual neighboring relationships, used to build a specific overlay graph. The overlay graph properties are thus exploited by the overlay routing protocol. Yet neighborhood in classical overlay networks is a virtual concept. Usually, there is no physical link between two nodes that are neighbor in the overlay network. Rather the two nodes are remote within the underlying physical topology, and the physical multi-hop path between the two nodes does not exhibit any particular graph property. This -- unless assuming the overlay network provided with a complete knowledge about the topology of the portion of the network where the two neighbor nodes belongs, which is unreasonable from a scalability perspective -- implies that the actual packet forwarding through the underlying physical network introduces a performance degradation that grows with the network size.

Conversely, entanglement-enabled overlay networks provide the nodes with entanglement-enabled links (\textit{e-link}) that can be used on-the-fly, without introducing any delay nor any performance degradation due to the mismatch between overlay and underlay network as in the classical case. Indeed, any quantum\footnote{As regards to classical signaling, it can be delayed further in time thanks to the deferred measurement principle.} overhead induced by the establishment of the e-links toward remote nodes occurred beforehand. Thus, the set-up process can be properly engineered for establishment e-links proactively as discussed in Section~\ref{Sec:3.3}, so that the actual entanglement utilization does not incur in any additional overhead.
 
\begin{figure*}
    \centering
    \begin{subfigure}[b]{0.8\textwidth}
        \includegraphics[width=1\linewidth]{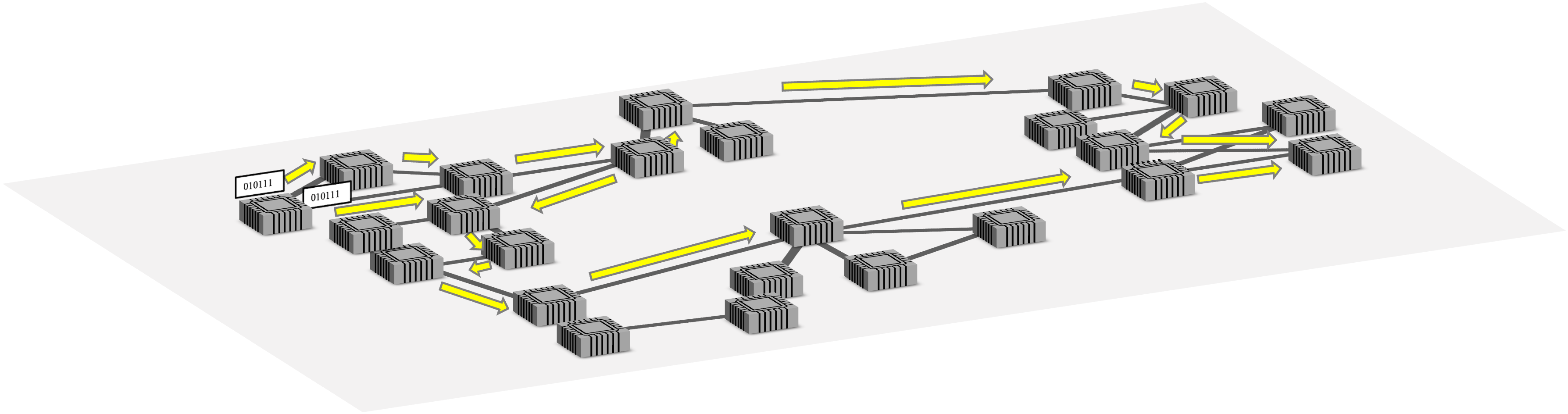}
        \caption{Classical multi-path routing: multiple copies of the same classical packet are forwarded through different classical paths (yellow arrows) and transmitted to the receiver.}
        \label{fig:Fig-03.a}
    \end{subfigure}
    \hfill
    \begin{subfigure}[b]{0.8\textwidth}
        \includegraphics[width=1\linewidth]{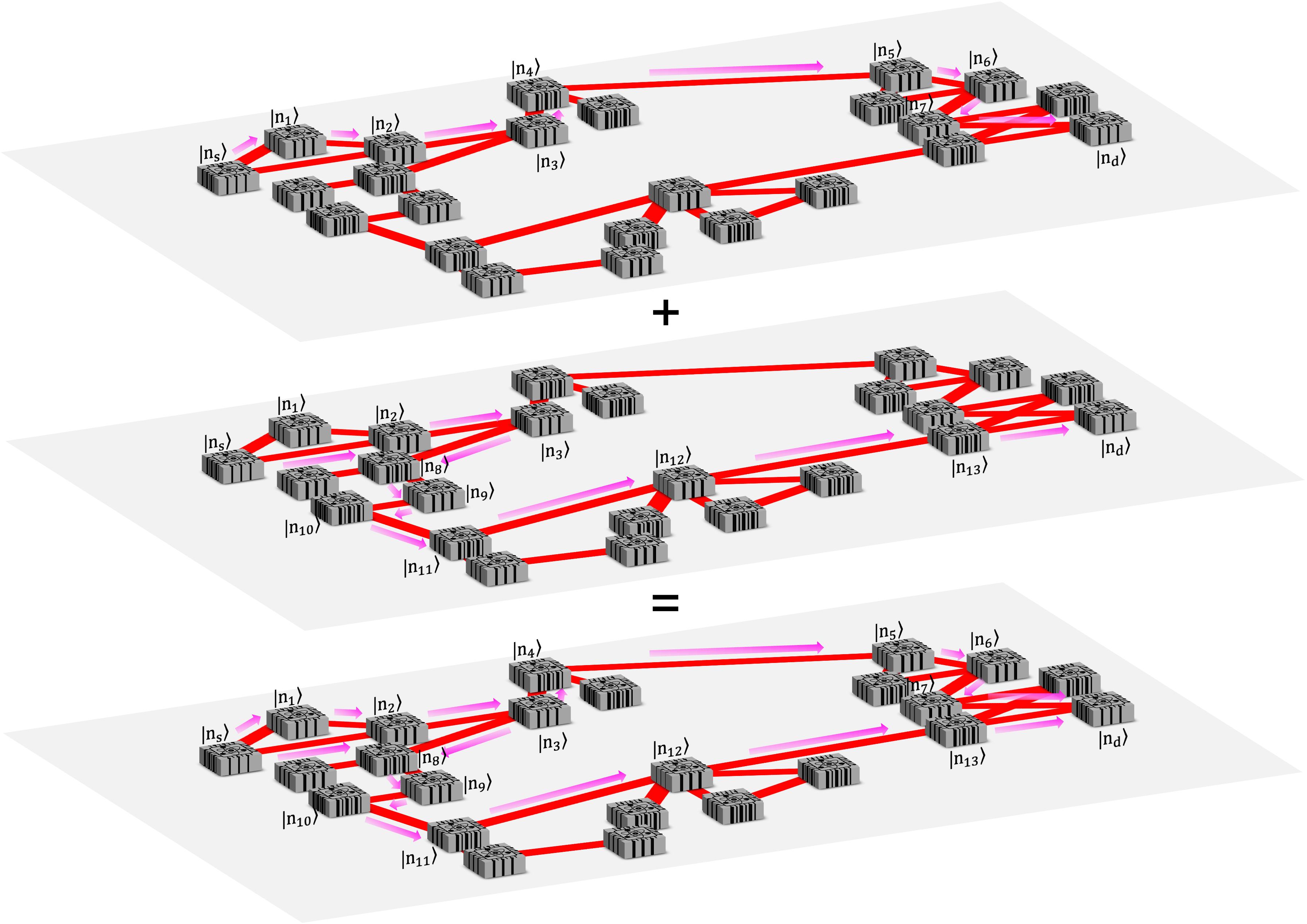}
        \caption{Quantum path: a quantum superposition of multiple paths allows the transmission of a single quantum packet simultaneously through them, without the violation of no-cloning theorem.}
        \label{fig:Fig-03.b}
    \end{subfigure}
    \caption{A pictorial representation of the concept of quantum path.}
    \label{fig:03}
    \hrulefill
\end{figure*}

% ------------------------------------------------------
\subsection{Quantum Path}
\label{Sec:3.2}

Existing models for the Quantum Internet protocol stack overlook an additional level of quantization, that comes into play when the unique propagation characteristics of quantum information carriers are taken into account. Specifically, counter-intuitively quantum mechanics allows a particle to propagate simultaneously among multiple space-time trajectories \cite{ChiKri-18,RamPirDur-21,CalSimCac-23}. This peculiar property enables scenarios where quantum information carrier propagates through a \textit{quantum path}, i.e., through a path\footnote{It is worthwhile to note that, despite their counter-intuitive nature, quantum paths have been already experimentally implemented, and they have been shown to provide significant advantages for a number of problems arising in both quantum computation and quantum communications, ranging from noise suppression to entanglement generation and distribution. We refer the reader to \cite{KouCacSim-22} for an in-depth overview of quantum paths.} in a quantum superposition of different configurations. This yields different powerful setups \cite{ChiKri-18}, such as superposition of different (in space) links or superposition of different alternative orders among the links. Accordingly, the communication path is 
quantized \cite{CaccIllKou-22}.

It is a matter of fact that the exploitation of a quantum path cannot rely on classical node addressing, which fails to capture the quantum features of the quantum paths. Specifically, once a quantum packet is sent on a quantum path, the ``packet location'' is not univocally determined since it is in superposition of different time/space configurations. Rather, the packet location is indefinite and, hence,  a quantum address is mandatory for describing such a superposition.   As instance, as depicted in Figure.~\ref{fig:Fig-03.b} a quantum route exploiting quantum paths can be implemented as superposition of paths or a superposition of orders \cite{RamPirDur-21,CaccIllKou-22}.  

The quantum path framework is a very powerful tool, key for any routing protocol genuinely quantum \cite{RamPirDur-21}, since it allows to significantly enhance the performance of the quantum network, by exploiting end-to-end paths with no-classical counterpart. Indeed, through a quantum path, the quantum carrier is delivered via different sets of intermediate nodes and different set of point-to-point links that exhibit different qualities of service. Hence, the genuine quantumness exhibited by the quantum path can exploit all the degrees of diversity (ranging from spatial through causal to temporal diversity), without any violation of the no-cloning theorem as it would happen by trying to adopt classical multi-path routing strategies \cite{CaccIllKou-22}. This is depicted in Figure~\ref{fig:03}.

Hence, for a successful quantum protocol stack design, it is key to recognize that providing the network nodes with a quantum address is mandatory for taking full advantage of the unique propagation characteristics of quantum carriers. As instance, with respect to the mentioned figure, the quantum packet propagates through a quantum route where the first hop is an even superposition of two quantum links, i.e., $e_{\ket{n_s},\ket{n_1}}$ and $e_{\ket{n_s},\ket{n_2}}$.

% ------------------------------------------------------
\subsection{Quantum Routing}
\label{Sec:3.3}

Up to now, existing literature on the Quantum Internet has considered quantum routing as the problem of distributing end-to-end entanglement between remote network nodes, according to some routing metric. As instance, several proposals have accounted for the temporal constraints induced by decoherence effects within the routing process, either by defining coherence-times-aware routing metrics or by incorporating these temporal constraints within the routing protocols. Similarly, several proposals have focused on optimizing entanglement distribution, with proposals ranging from fidelity maximization through purification/distillation to end-to-end path discovery. And a widely investigated area is constituted by the adoption of quantum repeaters, which combine entanglement swapping and entanglement purification to extend the entanglement over end-to-end path.

Yet, when it comes to quantum routing, there exists a fundamental difference with respect to classical routing that has been mainly overlooked so far.

Classical information is generated at the source for a given (usually, unicast) destination. Accordingly, classical routing goal is to \textit{find the ``best'' route toward the destination}, and indeed any classical routing metric measures the utility of a neighbor node in terms of its ``proximity'' toward the destination.

Conversely, the goal of the Quantum Internet routing is no longer to discover the route toward the destination. Rather, the goal is to \textit{entangle the source with the ``closest'' node that is already entangled with the destination}. 

At a first sight, classical and quantum routing goals might seem identical. In fact, someone might object that after all, for entangling these intermediate nodes with the destination, the very same problem underlying  classical routing -- i.e., discovering a path (from these intermediate nodes rather than from the source) toward the destination -- must have been solved beforehand.

But this objection overlooks a fundamental difference\footnote{The interested reader is referred to \cite{IllCalMan-22} for an in-depth treatise of the differences arising with quantum information and entanglement with respect to classical information.} between information and entanglement. Information, both classical and quantum, is valuable for the destination only. Any other intermediate node -- while forwarding it to the destination -- cannot exploit it for its communication needs. Hence, the beneficiary of classical and quantum information is fixed and pre-determined. Conversely, entanglement represents a communication resource valuable for any cluster of nodes sharing it, regardless of where it has been originally generated and regardless of the identities of the nodes originally supposed to use it. Indeed, the only requirement for exploiting an entangled qubit locally available is to coordinate with the other nodes sharing the entanglement resource. In a nutshell, while information exhibits a local, predetermined value, entanglement is characterized by a global, dynamic usefulness.

\begin{rem}
From the above, it follows that, whenever a \textit{proactive}\footnote{By borrowing ad-hoc networks terminology, we can classify the strategies for the entanglement distribution from a network engineering prospective as either \textit{proactive} or \textit{reactive}. Proactive strategies aim at early distribution of entanglement resources -- ideally, with a new generation process starting as soon as the entanglement resource is depleted -- whereas reactive strategies aim at on-the-fly distribution of entanglement, with a new generation process starting on demand, when needed.} entanglement distribution strategy is adopted, the Quantum Internet can exploit the additional degree of freedom represented by the global and dynamic usefulness exhibited by entanglement for providing the communication services, as illustrated in Section~\ref{Sec:4}.
\end{rem}

From the above, it becomes evident that classical location-aware addressing -- where routing tables store partial information toward clusters of destinations as it happens with classical IP -- fails in providing useful topological information activated by entanglement-based networks. Indeed, the overall objective of routing tables should switch from tracking next hops toward destinations to track entanglement resources.

With respect to this aspect, it is important to underline that entanglement is not limited to bipartite entangled states such as EPR pairs. Rather \textit{multipartite} entanglement greatly enriches the features of entanglement-based connectivity \cite{IllCalMan-22}, which in turn is deeply affected also by the specific properties characterizing the selected multipartite entanglement class\footnote{As instance, GHZ states constitute the natural substrate for applications aiming at distributively achieving some consensus or some form of synchronization, whereas W states represent a valuable tool for breaking any symmetry among the different parties, hence enabling applications based on leader election or distributed resource access \cite{IllCalMan-22}.}. It must be noted, though, that multipartite entanglement requires further coordination and signaling among the entangled nodes, when compared to EPR pairs. For this, network nodes aim at exploiting multipartite entangled state must be provided with the identities of all the nodes sharing such a state, along with the class of entanglement to whom the state belongs to.

\begin{rem}
Clearly, proactive entanglement distribution requires coherence times longer than those associated with the execution of the network functionalities. And, whenever this requirement cannot be satisfied, reactive entanglement distribution represents the only possible strategy.
\end{rem}

% ------------------------------------------------------
% Sec. IV
% ------------------------------------------------------
 
\section{From Software Defined Networks to Entanglement Defined Networks}
\label{Sec:4}
 
\begin{figure*}
    \centering
    \includegraphics[width=0.8\linewidth]{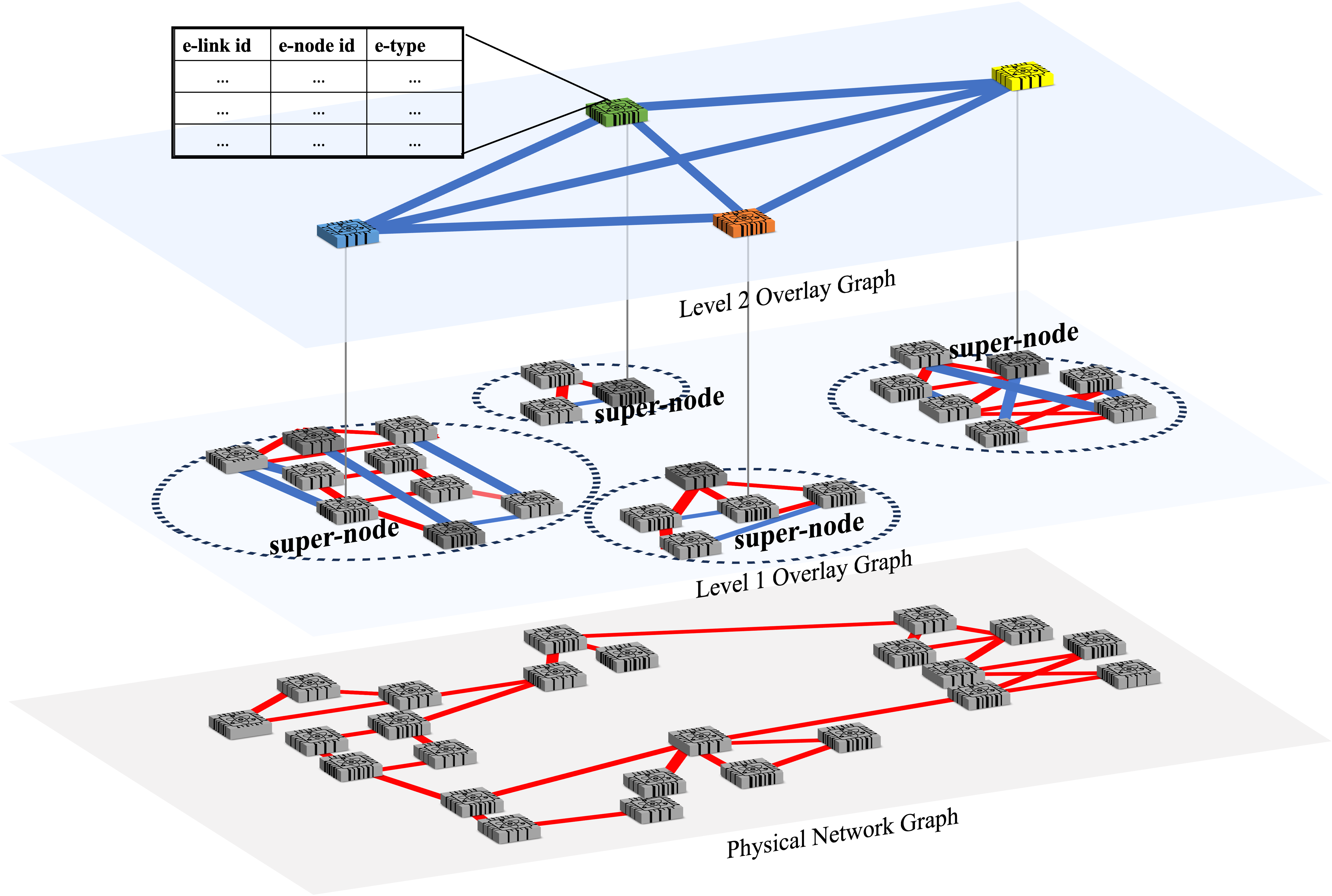}
    \caption{Toy-model representation of hierarchical quantum addressing, organized in a two-level hierarchy. Level-one clusters are organized with a single super-node serving some end-nodes, whereas level-two clusters are organized in a peer-to-peer topology among the super-nodes with augmented fully connectivity.}
    \label{fig:04}
    \hrulefill
\end{figure*}

% ------------------------------------------------------
The main idea for exploiting quantum addressing is to challenge the paradigm underlying classical hierarchical routing. Rather than designing a (classical) addressing scheme that enables scalable routing tables at the price of link depletion in the overlay routing network, we aim at designing a quantum addressing scheme that builds an overlay entangled network through link augmentation.

Specifically, in Figure~\ref{fig:04} we sketch a toy-model in which hierarchical principles are hybridized with entanglement marvels. Accordingly, within the figure, the nodes are organized in a two-level hierarchy. Level-one clusters are organized with a single super-node serving some end-nodes, whereas level-two clusters are organized in a peer-to-peer topology among the super-nodes with augmented fully connectivity. 

When it comes to the generation of entangled states, it is very reasonable, given the current maturity of quantum technologies, to assume a specialized super-node responsible for entanglement generation. The rationale for this assumption is twofold. On one hand, it accounts for the complex mechanisms and the dedicated equipment underlying the entanglement generation. On the other hand, it accounts for the mandatory requirement of some sort of local interaction among the qubits to be entangled. Accordingly, we consider the hierarchical overlay network in Figure~\ref{fig:04}. 

Clearly, the choice of the overlay entangled network in the figure is not either restrictive or univocal, since there exists two degrees of freedom in designing the overlay network that can (and should) be jointly optimized: i) clustering, and ii) connectivity within each cluster level.

With reference to clustering, although it is a complete new research area, we could envision to borrow some well-practices developed in the classical networking, by exploiting some physical topological information.

With reference, instead, to the augmented connectivity of level-two overlay graph, we highlight that the architecture of the entanglement-enabled connectivity plays a crucial role, since it determines the features of the level-two overlay network, and its capability to activate specific network functionalities. As a consequence, it should be recognized that the specific entanglement class(es) selected to realize the level-two overlay network is a design choice, which has to be carefully individuated. In this context, for instance, the amount of communication qubits available at each super-node to be devoted to maintain proactively the level-two overlay graph plays a crucial role, and it deeply influences the overall routing performance of the scheme built upon it. Preliminary research seems suggesting that memory and communications costs for augmented connectivity scale efficiently with the network size \cite{SchManTan-16,BenHajVan-23}, but this research area constitutes still an open issue.  

Furthermore, we note that the above mentioned design choices will influences the particulars of the interactions between the different overlay layers. In the vision developed through Figure~\ref{fig:04}, we envision that the level-two hierarchy is also responsible for maintaining topological information needed to navigate the overlay graph and to fulfill the communication needs of the end-nodes. Specifically, as discussed in Section~\ref{Sec:4}, quantum routing requires a paradigm shift with respect to classical routing. This becomes particularly evident by inspecting the information stored within the routing tables in Figure~\ref{fig:04}.
Quantum routing communication opportunities are not represented as a classical link interface toward a (physical) next hop, but they are rather represented as an entanglement interface -- namely, one or more communication qubits stored within the node -- toward a neighbor node within the overlay entangled network. As a matter of fact, multipartite entanglement requires additional information -- such as the identities of all the nodes (\textit{e-node} column of table in Figure~\ref{fig:04}) sharing the resource and the particular class to whom the entangled resource belongs (\textit{e-type} column of table in Figure~\ref{fig:04}) -- to be stored within the table, as shown in Figure~\ref{fig:04}.

\vspace{5mm}
% ------------------------------------------------------

 {\begin{rem}
   By maintaining proactively the level-two overlay graph, the very concept of quantum routing is changed. Indeed, the quantum routing problem can be efficiently solved via quantum algorithms, which exploit the entanglement-based overlay graph and the routing tables available at the super-nodes. Preliminary research about distributed Grover algorithms goes in this direction \cite{ChaNovAmb-16}. However, further research is needed to design effective quantum algorithms able to exploit the quantum path concept and the entanglement-based overlay graph. 
\end{rem}}

Furthermore, level-two overlay graph could be further exploited to coherently control, through the quantum addresses, the involved super-nodes so that entanglement is generated without the need of physically navigating the graph, as suggested preliminary in \cite{KouCacCal-23}, by exploiting the quantum path framework. 

From the above, we are building a new Quantum Internet ecosystem, which moves from the software-defined paradigm to the \textit{entanglement-defined one}.

% ------------------------------------------------------------------------------------
% Sec. V
% ------------------------------------------------------------------------------------
 
\section{Conclusion and Open Issues}

Short-term efforts toward Quantum Internet are reasonably trying to reconcile quantum information and quantum entanglement to classical information -- with an approach that can be defined as \textit{design by analogy} \cite{DiaQiMil-22}. This research activity -- although incremental from a network design perspective -- is unquestionably important for the deployment of pilot small-scale networks, as well as from telcom operators' viewpoint aiming at maximizing current network 
assets revenue.

Yet, from disruptive, long-term perspective, quantumness and its unconventional features should not be overlooked. Rather, they should be spotlighted and emphasized to have a deep impact on the network design, radically influencing the quantum network functionalities through a major paradigm shift, somehow similar to the shift from circuit-switching to packet-switching design for classical networks \cite{IllCalMan-22}. 

In the light of disruptive, long-term perspective, with this manuscript we aimed at highlighting the quest for a genuine Quantum Internet Protocol, which is currently missing.

We believe that the Quantum Internet Addressing design should came from a collaborative standardization effort, and we do hope that this manuscript can fuel the starting of this process, which requires to address several key research issues, as described in details within the manuscript and briefly summarized in the following:
\begin{itemize}
    \item [i)] the design of a quantum addressing scheme able to capture the dynamic nature of the entanglement-enabled connectivity;
    \item [ii)] the engineering of the quantum addressing scheme for effectively exploiting quantum paths;
    \item [iii)] the design and engineer an \textit{overlay entangled network} where the network graph properties needed by the routing process are obtained through \textit{topological augmentation}, rather then topological depletion; this design should also properly define the particulars of the interactions between the different overlay layers;
    \item [iv)] the choice and optimization of the specific entanglement class(es) selected to realize the hierarchical levels of the overlay graph;
    \item [v)] the design of quantum algorithms able to efficiently solve quantum routing problems by exploiting the quantum addressing scheme;
    \item[vi)] the availability of quantitative performance evaluation tools and network simulators, targeting genuine Quantum Internet functionalities; 
\end{itemize}

Although we have more doubts than answers, we do look forward to contribute to such an exciting research area, which could pave the way for the Internet of future such as Arpanet paved the way for today’s internet.

% --------------------------------------------------------------------------
% Ack
% --------------------------------------------------------------------------
\section*{Acknowledgment}

The authors thank Michele Viscardi for proofreading the early version of this manuscript and for helping in drawing Figure~\ref{fig:03}.

% --------------------------------------------------------------------------
% Biblio
% --------------------------------------------------------------------------
\bibliographystyle{IEEEtran}
\bibliography{biblio.bib}

\end{document}